\documentclass[11pt]{elsarticle}
\usepackage{gensymb}
\usepackage{mathtools}
\usepackage{amsfonts}
\usepackage{amssymb}
\usepackage{amsopn}
\usepackage{amsmath,amssymb,amsfonts,bm,amsthm,latexsym} 
\usepackage[utf8]{inputenc}

\pdfoutput=1

\makeatletter
\def\ps@pprintTitle{%
  \let\@oddhead\@empty
  \let\@evenhead\@empty
  \let\@oddfoot\@empty
  \let\@evenfoot\@oddfoot
}
\makeatother

\usepackage{url}
\usepackage{breakurl}
\usepackage[breaklinks,
            colorlinks = true,
            linkcolor = blue,
            urlcolor  = blue,
            citecolor = blue,
            anchorcolor = blue]{hyperref}

\usepackage{lineno}

\usepackage{graphicx}
\usepackage[margin=1.25in]{geometry}
\usepackage[usenames,dvipsnames]{color}
\newcommand\acp{\begin{center}
\rule[-0.2in]{\hsize}{0.01in}\\\rule{\hsize}{0.01in}\\
\vskip 0.1in Submitted to the  Proceedings\\ 
of the African Conference on Fundamental and Applied Physics
    \vskip 0.05in
    {\it Second Edition, ACP2021, March 7--11, 2022 --- Virtual Event}\\
\rule{\hsize}{0.01in}\\\rule[+0.2in]{\hsize}{0.01in} \\
\end{center}}

\usepackage[firstpage=true]{background}
\backgroundsetup{contents={\parbox{6.5in}{\acp}}, scale=1,placement=top,opacity=1,color=black,position={3.25in,1.2in}}

\usepackage{fancyhdr}
\fancypagestyle{plain}{%
  \fancyhf{}%
  \fancyhead[C]{}
  \fancyfoot[C]{\thepage}
}

\fancypagestyle{empty}{%
  \fancyhf{}%
  \fancyhead[C]{{\it ACP2021, March 7--11, 2022 --- Virtual Edition}}
  \fancyfoot[C]{\thepage}
}
\def\div {\mbox{div}\,}

\def\ra{\rangle}
\def\la {\langle}

\def\ab{\al\bt}

\newcommand{\ben}{\begin{enumerate}}
\newcommand{\een}{\end{enumerate}}

\newcommand{\bef}{\begin{frame}}
\newcommand{\eef}{\end{frame}}
\newcommand{\bi}{\begin{itemize}}
\newcommand{\ei}{\end{itemize}}
\newcommand{\itt}{\item}
\newcommand{\bfl}{\begin{flushleft}}
\newcommand{\efl}{\end{flushleft}}

\newcommand{\bb}{\begin{block}}
\newcommand{\eb}{\end{block}}

\newcommand{\ep}{\epsilon}

\newcommand{\be}{\begin{equation}}
\newcommand{\bse}{\begin{subequation}}
\newcommand{\ese}{\end{subequation}}
\newcommand{\ee}{\end{equation}} %\indent}
\newcommand{\eei}{\end{eqnarray}\indent\indent}
\newcommand{\bc}{\begin{center}}
\newcommand{\ec}{\end{center}}
\newcommand{\ber}{\begin{eqnarray}}
\newcommand{\eer}{\end{eqnarray}}
\newcommand{\bern}{\begin{eqnarray*}}
\newcommand{\eern}{\end{eqnarray*}}
\newcommand{\beast}{\begin{equation*}}
\newcommand{\eeast}{\end{equation*}}
\newcommand{\ba}{\begin{array}}
\newcommand{\ea}{\end{array}}
\newcommand{\bal}{\begin{align}}
\newcommand{\eal}{\end{align}} 
 
\newcommand \om {\omega}

\newcommand{\D}{\tl\nb}
\newcommand{\al}{\alpha}
\newcommand{\bt}{\beta}

\newcommand{\sfrac}[2]{{\textstyle{#1\over#2}}}
\def\case#1/#2{\textstyle\frac{#1}{#2} }
\newcommand{\nb}{\nabla}

\newcommand{\tl}{\tilde}
\newcommand{\bver}{\begin{verbatim}}
\newcommand{\ever}{\end{verbatim}}

\newcommand{\de}{\delta}
\newcommand{\lam}{\lambda}

\newcommand{\Tht}{\Theta}
\newcommand{\gam}{\gamma}

\newcommand{\gad}{\gamma\delta}
\newcommand{\abg}{\alpha\beta\gamma}

\def\ab{\al\bt}

\begin{document}

\begin{frontmatter}

%\pubblock

\title{Dark-fluid constraints of shear-free universes}

\author{Amare Abebe}
\ead{AmareAbebe.Gidelew@nwu.ac.za}

%\cortext[cor1]{Corresponding Author}

\address{Center for Space Research, North-West University, Mahikeng 2745, South Africa}

\begin{abstract}
\noindent 
 Integrability conditions arising from general irrotational fluid-flow considerations of a universe dominated by cosmic dark fluids will be investigated under special assumptions on the nature of the spacetime shear.  Special emphasis will be placed on linearized perturbations of quasi-Newtonian and anti-Newtonian spacetimes, whereby the conditions for the existence and consistent evolution of such spacetimes in the presence of the Chaplygin gas fluid model will be derived and discussed. 
\end{abstract}


\end{frontmatter}

%\linenumbers
%

%\def\thefootnote{\fnsymbol{footnote}}
%\setcounter{footnote}{0}

%\newpage

\section{Introduction}
\label{sec:intro}
\noindent
The recent discovery of the accelerated rate of cosmic expansion has inspired a wave of new research into the nature of gravitational physics. As a result, new alternatives to the solutions of General Relativity (GR) theory abound, and dark-fluid (DF) models are among those alternatives: aimed at addressing dark matter and dark energy issues in a unified framework \cite{kamen01,elmardi17}. The Chaplygin gas (CG) model is among the most widely explored DF (see \cite{elmardi17} and references therein).

\noindent  In order to understand the dynamics of nonlinear fluid flows in cosmology, it is important to understand the relationship between their Newtonian and GR limits, and this usually necessitates studying the differential properties of time-like geodesics that describe the fluid flow. The expansion $\Theta$, shear $\sigma_{\ab}$, rotation $\omega^\al$, and acceleration $A_a$ of the four-velocity field $u^a$ tangent to the fluid flowlines describe the kinematics of such fluid flows.

\noindent The main objective of this work is to make a consistency analysis of the field equations for models with vanishing shear $\sigma_{\ab}$ in CG-dominated universes, but generally non-vanishing energy density $\mu\;,$ vorticity $\om_\al$ and a locally free gravitational field covariantly described by the gravito-electric (GE) and gravito-magnetic (GM) components of the Weyl tensor, $E_{\ab}$ and $H_{\ab}\;$ respectively.

\noindent The rest of this work is organized as follows: the covariant description of the background spacetime and the cosmological field equations, together with their limiting solutions, are presented in Sec.~\ref{sec:cov}, followed by a conclusion in Sec.~\ref{sec:conc}.

\section{Covariant description of shear-free spacetimes with a dark fluid}
\label{sec:cov}
\noindent
The standard GR  field equations  are given (in units where $8\pi G=1=c$) by
 \beast
 G_{\al\bt}= T_{\al\bt}\;, \quad T_{\al\bt} \equiv \mu u_{\al}u_{\bt} + ph_{\al\bt}+ q_{\al}u_{\bt}+ q_{\bt}u_{\al}+\pi_{\al\bt}\;,
 \eeast
where the first (geometric) term $G_{\ab}$ is represented by the Einstein tensor, and the energy-momentum tensor of matter fields $T_{\ab}$ is the source of $\mu$, $p$, $q_{a}$ and $\pi_{\al\bt}$, i.e.,  the energy density, the isotropic
pressure, the heat flux and the anisotropic pressure of the fluid, respectively.  $u^{\al}$ is the $4$-velocity of fundamental observers comoving with the fluid.

\noindent We assume that the universe contains the CG as one of the component fluids, the other components being dust, radiation, etc.  In such a multi-component cosmic medium, the total energy density, the isotropic and anisotropic pressures, and the heat flux are given, respectively, by
\beast\label{multif}
\mu\equiv\sum_i \mu^{i}\quad\quad\quad\;p\equiv \sum_i p^i\quad\quad\quad q_{\al}\equiv \sum_iq^{i}_{\al}\quad\quad\quad\pi_{\ab}\equiv\sum_i \pi^{i}_{\ab}
\eeast
with the index $i$ labelling the thermodynamic property of the $i^{th}$ fluid.
 If we assume the late time matter distribution to be dominated by dust and the CG, then we can write:
\beast\label{totaltherm}
\mu=\mu^{d}+\mu^{c}\quad\quad\quad\;p=p^{c}\quad\quad\quad q_{\al}= q^{d}_{\al}+q^{c}_{\al}\;,
\eeast
 where  $p^d=0$ and $ \pi^d_{\ab}=0=\pi^c_{\ab}$ are generally assumed. We further assume the normalized 4-velocity $u_\al$ of fundamental observers coincides with that of standard matter $u^d_\al$ such that
\beast\label{frames}
v^c_\al\equiv u^c_\al-u_\al\quad\quad q^d_\al=0\quad\quad q^c_\al=(\mu^c+p^c)v^c_\al
\eeast
where the CG's 4-velocity is tilted w.r.t to $u_\al$ by the peculiar velocity $v^c_\al\ll1$ (a non-relativistic approximation).
 We need the following evolution and their constraints ($C^1$ - $C^4$) for our consistency analysis \cite{abebe21, maartens98}:
\ber
\label{qec}\dot{q}^{c}_{\al}&=-&\sfrac{4}{3}\Theta q^{c}_{\al}-(\mu^c+p^c)A_\al-\D_{\al}p^{c}\\
\label{propom}\dot{\omega}_{\al}&=&-\sfrac23\Theta\omega_{\al}-\sfrac{1}{2}\ep_{\abg}\tl\nb^{\bt}A^{\gam}\\
\label{sig}\dot{\sigma}_{\ab}&=&-\sfrac{2}{3}\Theta\sigma_{\ab}-E_{\ab}+\tl\nb_{\la \al}A_{\bt\ra}\\
\label{gep}\dot{E}_{\ab}&=&\ep_{\gad\langle \al}\tl\nb^{\gam}H^\de{}_{\bt\rangle }-\Theta E_{\ab}-\sfrac{1}{2}\left(\mu+p\right)\sigma_{\ab}
-\sfrac{1}{2}\tl\nb_{\langle \al}q^c_{\bt\rangle}\\
\label{shearcon} (C^{1})_{\al}&:=&\D^{\bt}\sigma_{\ab}-\sfrac{2}{3}\tl\nb_{\al}\Theta+\ep_{\abg}\tl\nb^{\bt}\omega^{\gam}+q^c_{\al}=0\\
\label{omcon} (C^{2})_{ \ab}&:=&\ep_{\gad(\al}\tl\nb^{\gam}\sigma^\de{}_{\bt)}+\tl\nb_{\langle \al}\omega_{\bt \rangle}-H_{\ab}=0\\
\label{hcon} (C^{3})_{\al}&:=&\tl\nb^{\bt}H_{\ab}+(\mu+p^c)\omega_{\al}+\sfrac12\ep_{\ab\de}\tl\nb^{\bt}q^{\de}_c=0\\
\label{econ} (C^{4})_{\al}&:=&\tl\nb^{b}E_{\ab}-\sfrac13\tl\nb_{\al}\mu+\sfrac13\Theta q^c_{\al}=0
\eer

\noindent Over the years, the role of shear in GR and the special nature of shear-free cases in particular have been studied. 
 G\"{o}del showed \cite{godel52} that shear-free time-like geodesics of some spatially homogeneous universes cannot expand and rotate simultaneously and this result was later generalized \cite{ellis67,nzioki11,ritu21} to include inhomogeneous cases of shear-free time-like geodesics.
 Let us investigate the effect of switching off the shear term from the above evolution and constraint equations relevant for our present work.  A first observation is that Eq. \eqref{sig} turns into a new constraint equation:
\be
\label{shearconst2} (C^{6})_{\ab}:=E_{\ab}-\tl\nb_{\la \al}A_{\bt\ra}=0\;.
\ee
 The special case where $q^c_\al=0$ in Eq. \eqref{qec} turns into a further constraint 
\be\label{A1}
A_\al=-\frac{\D_{\al}p^{c}}{\mu^c+p^c}
\ee was recently investigated \cite{abebe21} and shown to have counter-examples to the generalized Ellis shear-free conjecture.
Simultaneously expanding ($\Theta\neq 0$) and rotating ($\om^a\neq 0$) fluid flow solutions exist in CG-dominated cosmological models.
 These counter-examples, however,  force a special algebraic relationship between the defining CG fluid parameters.
Beyond these counter-examples, any expanding shear-free CG-dominated universe with vanishing heat flux  must generally be non-rotating.
 A direct implication of this will be that from Eq. \eqref{propom} another new constraint emerges:
\be\label{A2}
\ep_{\abg}\tl\nb^{\bt}A^{\gam}=0\implies A_\al=\tl\nb_\al\phi
\ee
for some scalar potential $\phi$. Comparing Eqs. \eqref{A1} and \eqref{A2}, one concludes
\be
\phi=-\frac{1}{2}\ln\left(\frac{\mu^c+p^c}{\mu^c}\right)=-\frac{1}{2}\ln\left(1-\frac{A}{\mu^2_c}\right)\;.
\ee
An even more interesting consequence of the above special case will be that, due to Eq. \eqref{omcon}, $H_{\ab}$ identically vanishes, leading to the so-called {\it quasi-Newtonian} universe with a homogeneous expansion, since $\tl\nb_\al\Tht=0$ in Eq. \eqref{shearcon}. Such models are generally unstable to linear perturbations and do not support large-scale structure formation because
\be
\tl\nb_\alpha\Tht=0\implies \tl\nb_\al\mu=0\implies \tl\nb_\al\mu^d+\tl\nb_\al\mu^c=0\;.
\ee
This shows that there has to be a fine balance between dust and the CG such that any tendency for structures to grow out of dust perturbations will be discouraged by those of the latter. Let us now consider shear-free models with a net heat-flux due to the CG fluid, while limiting our focus to irrotational-fluid cases for now. An immediate consequence of the irrotational-fluid assumption would be that from Eq.  \eqref{hcon},
\be
\ep_{\ab\de}\tl\nb^{\bt}q^{\de}_c=0\implies q^c_\al=\tl\nb_\al\psi =\frac{2}{3}\tl\nb_\al\Tht\implies \psi=\frac{2}{3}\Tht+C
\ee
for some scalar potential $\psi$ and (spatial) constant $C$. We can then show that the peculiar velocity of the CG fluid (w.r.t the worldline of the fundamental observers) can be given, either in terms of the expansion gradient or total energy density gradient,  by
\be\label{velqn}
v^c_\al=\frac{2}{3(\mu^c+p^c)}\tl\nb_\al\Tht=\frac{1}{(\mu^c+p^c)\Tht}\tl\nb_\al\mu\;.
\ee
Using this result together with Eq. \eqref{qec} the acceleration of the fluid can be shown to be
\be\label{accqn}
A_\al=\dot{v}^c_\al+\left(\frac{1}{3}+\frac{A}{\mu^2_c}\right)\Tht v^c_\al+\frac{A}{\mu^2_c(\mu^c+p^c)}\tl\nb_\al\mu^c\;,
\ee
 and this generalizes the quasi-Newtonian relation obtained for dust \cite{maartens98}. 
 
 \noindent Another class of models we want to investigate are those with
$E_{\ab}=0$. Referred to as anti-Newtonian spacetimes, these models are considered to be the farthest from the Newtonian theory.
  Drawing parallels to the quasi-Newtonian case of vanishing $\sigma_{\ab}$ and $q^c_\al$,  one immediately observes that such spacetimes cannot be shear-free, for if we allow the shear to vanish, $H_{\ab}$ would have to vanish as well.  A vanishing heat flux leads to the Friedman-Lema\^itre-Robertson-Wlaker (FLRW) spacetime:
\be
\tl\nb_\al\Tht=0=\tl\nb_\al\mu\implies\mbox{~FLRW~ background ~spacetime}.
\ee
For $q^c_\al\neq 0$, we obtain similar forms of $q^c_\al\;, v^c_\al$ and $A_\al$ with the extra condition
\be
\tl\nb_{\langle \al}q^c_{\bt\rangle}=0\implies \tl\nb_{\langle \al}v^c_{\bt\rangle}=0=\tl\nb_{\langle \al}\tl\nb_{\bt\rangle}\mu\;.
\ee
Considering $E_{\ab}=0\;,\sigma_{\ab}\neq 0$ results in Eq. \eqref{gep} turning into a new constraint:
\be
\label{gec}(C^7)_{\ab}=\ep_{\gad\langle \al}\tl\nb^{\gam}H^\de{}_{\bt\rangle }-\sfrac{1}{2}\left(\mu+p\right)\sigma_{\ab}
-\sfrac{1}{2}\tl\nb_{\langle \al}q^c_{\bt\rangle}\;,
\ee
whereas Eqs. \eqref{hcon} and \eqref{econ}  lead to the conclusion that:
\be\label{qantin}
\tl\nb_{\al}\mu=\Theta q^c_{\al}\implies \tl\nb^{\bt}H_{\ab}=0\;.
\ee
A necessary condition for the propagation of gravitational radiation is the vanishing of the divergence of a non-vanishing $H_{\ab}$. Eq. \eqref{qantin} therefore shows that gravitational radiation can propagate in a CG-dominated anti-Newtonian universe, as opposed to the [quasi-]Newtonian solutions.  Using Eqs. \eqref{shearcon}, \eqref{qantin} and $\Tht^2=3\mu$, we see that for the anti-Newtonian model
\be
q^c_\al=\sfrac{2}{3}\tl\nb_{\al}\Theta-\D^{\bt}\sigma_{\ab}\implies \tl\nb^{\bt}\sigma_{\ab}=0\;.
\ee
 The solutions for $q^c_\al\;, v^c_\al$ and $A_\al$ that we found in the quasi-Newtonian limit retain their forms in this class of models as well, subject to the constraint Eq. \eqref{gec}.

\section{Conclusions}\label{sec:conc}
In this work, we have seen the spacetime properties of the universe containing the CG model as a possible DF alternative.  We have shown the constrained evolutions of the shear-free spacetimes with quasi-Newtonian and anti-Newtonian solutions. Possible new frontiers in this direction include the analysis of the density and velocity perturbations and their nonlinear generalizations, considering more generalized CG models, and comparing the results with existing and future observational data.

%\section*{Acknowledgments}
% I acknowledge that this work is based on the research supported in part by the National Research Foundation (NRF) of South Africa (grant number 112131).
%\section*{References}
\bibliographystyle{elsarticle-num}
\bibliography{myreferences} 

\section*{Appendix}
The 4-velocity vector  $u^{\al}\equiv \frac{dx^{\al}}{dt}$ is used to define the covariant time derivative for any tensor 
${S}^{\al..\bt}{}_{\gam..\de} $ along an observer's worldlines:
\be
\dot{S}^{\al..\bt}{}_{\gam..\de}{} = u^{\lam} \nb_{\lam} {S}^{\al..\bt}{}_{\gam..\de}\;.
\ee
The projection tensor into the tangent 3-spaces orthogonal to $u^\al$ is given by
 \be h_{\al\bt}\equiv g_{\ab}+u_\al u_\bt
 \ee
and is used to define the fully orthogonally 
projected covariant derivative for any tensor ${S}^{\al..\bt}{}_{\gam..\de} $:
\be
\tl\nb_{\lam}S^{\al..\bt}{}_{\gam..\de}{} = h^{\al}{}_\mu h^\nu{}_\gam...h^{\bt}{}_\theta h^\phi{}_\delta 
h^\tau{}_\lam \nb_{\tau} {S}^{\mu..\theta}{}_{\nu..\phi}\;,
\ee
with total projection on all the free indices. The orthogonally projected symmetric trace-free part of vectors and rank-2 tensors is defined as
\be
V^{\langle \al \rangle} = h^{\al}{}_{\bt}V^{\bt}~, ~ S^{\langle \ab \rangle} = \left[ h^{(\al}{}_\gam {} h^{\bt)}{}_\de 
- \sfrac{1}{3} h^{\ab}h_{\gam\de}\right] S^{\gam\de}\;,
\label{PSTF}
\ee
and the volume element for the rest spaces orthogonal to $u^\al$ is given by 
\be
\ep_{\al\bt\gam}=u^{\de}\eta_{\de\al\bt\gam}=-\sqrt{|g|}\delta^0{}_{\left[ \al \right. }\delta^1{}_\bt\delta^2{}_\gam\delta^3{}_{\left. \de \right] }u^\de\Rightarrow \ep_{\al\bt\gam}=\ep_{[\al\bt\gam]},~\ep_{\al\bt\gam}u^{\gam}=0,
\ee
where $\eta_{abcd}$ is the 4-dimensional volume element with the properties
\be
\eta_{\al\bt\gam\de}=\eta_{[\al\bt\gam\de]}=2\ep_{\ab[\gam}u_{\de]}-2u_{[\al}\ep_{\bt]\gam\de}.
\ee
We define the covariant spatial divergence and curl of vectors and rank-2 tensors as
\ber
&& \div V=\D^\al V_\al\,,~~~~~~(\div S)_\al=\D^\bt S_{\ab}\,, \\
&& \text{curl }V_\al=\ep_{\al\bt\gam}\D^\bt V^\gam\,,~~ \text{curl} S_{\ab}=\ep_{\gam\de(\al}\D^\gam S_{\bt)}{}^\de \,.
\eer
The first covariant derivative of $u^a$ can be split into its
irreducible parts as
\be\label{congruence}
\nb_\al u_\bt=-A_\al u_\bt+\sfrac13\Theta h_{\ab}+\sigma_{\ab}+\ep_{\al \bt\gam}\omega^\gam,
\ee
where 
\be
A_\al\equiv \dot{u}_\al\;, \quad\Theta\equiv \tl\nb_\al u^\al\;,\quad 
\sigma_{\ab}\equiv \tl\nb_{\langle \al}u_{\bt \rangle}\;,\quad \omega^{\al}\equiv\ep^{\al \bt \gam}\tl\nb_\bt u_\gam\;.
\ee
Splitting the Weyl conformal curvature tensor 
\be
C^{\ab}{}_{\gam\de}\equiv R^{\ab}{}_{\gam\de}-2g^{[\al}{}_{[\gam}R^{\bt]}{}_{\de]}+\frac{R}{3}g^{[\al}{}_{[\gam}g^{\bt]}{}_{\de]}
\ee
 into its EM and GM  components, respectively, gives
\be
E_{\ab}\equiv C_{\al\gam\bt\de}u^{\gam}u^{\de},~~~~~~~H_{\ab}\equiv\sfrac{1}{2}\eta_{\al\theta}{}^{\gam\de}C_{\gam\de\bt\lam}u^{\theta}u^{\lam}\;.
\ee
\bi
\itt Complete set of linearised propagation equations:
\bern
\dot{\mu}^{d} &=&-\mu^{d}\Theta\\
\dot{\mu}^{c}&=&-(\mu^{c}+p^{c})\Theta-\D^{\al}q^{c}_{\al}\\
\dot{\Theta}&=&-\sfrac13 \Theta^2-\sfrac12(\mu+3p^c)+\tl\nb_\al A^\al\\
\dot{q}^{c}_{\al}&=&-\sfrac{4}{3}\Theta q^{c}_{\al}-(\mu^c+p^c)A_\al-\D_{\al}p^{c}\\
\dot{\omega}_{\al}&=&-\sfrac23\Theta\omega_{\al}-\sfrac{1}{2}\ep_{\abg}\tl\nb^{\bt}A^{\gam}\\
\dot{\sigma}_{\ab}&=&-\sfrac{2}{3}\Theta\sigma_{\ab}-E_{\ab}+\tl\nb_{\la \al}A_{\bt\ra}\\
\dot{E}_{\ab}&=&\ep_{\gad\langle \al}\tl\nb^{\gam}H^\de{}_{\bt\rangle }-\Theta E_{\ab}-\sfrac{1}{2}\left(\mu+p\right)\sigma_{\ab}
-\sfrac{1}{2}\tl\nb_{\langle \al}q^c_{\bt\rangle}\\
\dot{H}_{\ab} &=&-\Theta H_{\ab}-\ep_{\gad\langle \al}\tl\nb^{\gam}E^\de{}_{\bt\rangle }
\eern
\itt  Linearised constraint equations:
\bern
 (C^{1})_{\al}&:=&\D^{\bt}\sigma_{\ab}-\sfrac{2}{3}\tl\nb_{\al}\Theta+\ep_{\abg}\tl\nb^{\bt}\omega^{\gam}+q^c_{\al}=0\\
(C^{2})_{ \ab}&:=&\ep_{\gad(\al}\tl\nb^{\gam}\sigma^\de{}_{\bt)}+\tl\nb_{\langle \al}\omega_{\bt \rangle}-H_{\ab}=0\\
(C^{3})_{\al}&:=&\tl\nb^{\bt}H_{\ab}+(\mu+p^c)\omega_{\al}+\sfrac12\ep_{\ab\de}\tl\nb^{\bt}q^{\de}_c=0\\
(C^{4})_{\al}&:=&\tl\nb^{b}E_{\ab}-\sfrac13\tl\nb_{\al}\mu+\sfrac13\Theta q^c_{\al}=0\\
 \label{R5} (C^{5})&:=&\tl\nb^\al\omega_\al=0
\eern
\ei

\end{document}